# Optimized Distributed Processing in a Vehicular Cloud Architecture

**Fatemah S. Behbehani, Mohamed Musa, Taisir Elgorashi, and J. M. H. Elmirghani**
*School of Electronic and Electrical Engineering, University of Leeds, Leeds, LS2 9JT, United kingdom*

**ABSTRACT**

The introduction of cloud data centres has opened new possibilities for the storage and processing of data, augmenting the limited capabilities of peripheral devices. Large data centres tend to be located away from the end users which increases latency and power consumption in the interconnecting networks. These limitations led to the introduction of edge processing where small distributed data centres or fog units are located at the edge of the network close to the end user. Vehicles can have substantial processing capabilities, often un-used, in their on-board-units (OBUs). These can be used to augment the network edge processing capabilities. In this paper we extend our previous work and develop a mixed integer linear programming (MILP) formulation that optimizes the allocation of networking and processing resources to minimize power consumption. Our edge processing architecture includes vehicular processing nodes, edge processing and cloud infrastructure. Furthermore, in this paper our optimization formulation includes delay. Compared to power minimization, our new formulation reduces delay significantly, while resulting in a very limited increase in power consumption.

**Keywords**: Distributed Processing, Energy Consumption, Delay, Edge Nodes, Vehicular Networks,

## 1. INTROUCTION

Data centres and centeralized clouds have resulted in a significant shift in the capabilities of users in terms of data processing and data storage, thus significantly augmenting edge devices [1]. Traffic continues to grow at 30% - 40% [2] per year currently and this huge growth in traffic has led to a corresponding large growth in data centre requirements and in the power consumption of the network and processing [2]-[6]. Consideration has been given to different network segments [7] - [11], and to content distribution networks [12]. The introduction of distributed mini data centres close to the edge of the network is one promising solution that is currently being pursued [13] - [17]. Particular attention has been given to IoT [18], [19], the big data produced, and hence the improvement of the energy efficiency of such big data edge processing networks [20] – [23]. The introduction of autonomous driving promises new frontiers in vehicle processing capabilities, and hence the evaluation of Internet of Vehicle (IoV) approaches has gained momentum [24]. This added vehicular processing capabilities have opened new research avenues in vehicular clouds and in Vehicle as a Resource (VaaR) approaches [25]. We have considered the use of the processing capabilities of vehicles in [26] where vehicles become the first stage in the chain of processing options that extends in our architecture from the vehicle processors, to the edge layer and its potential fog processing and finally to the core network segment where the central clouds are located. We compared our architecture to conventional architectures where power consumption was the main criterion in our comparisons. Following the introduction, we present our vehicular distributed architecture in Section II. We then discuss our MILP model and the optimization in Section III together with the results. Finally, conclusions are given in Section IV.

## 2. VEHICULAR DISTRIBUTED COMPUTING ARCHITECTURE

Our proposed distributed processing architecture is shown in Fig. 1. There are several processing options in this architecture. Firstly, data can be processed in the processors available through the vehicle OBU, possibly augmented through a "processing box / server" fitted in the vehicle and connected to the vehicle CAN bus which also links to the vehicle OBU. This "processing box / server" ensures that the critical systems of the vehicle are isolated from external users, hence improving security. It can furthermore augment the OBU processing capabilities and can introduce uniform processing capabilities in different car models and aid rapid deployment. Here vehicular users who wish to participate and receive financial or credit rewards, can opt to have such a "processing box / server" fitted. Following the vehicle, processing can next be performed in servers placed close to users at the edge nodes. Communication between these edge nodes and the vehicles can be achieved through IEEE 802.11pp WiFi, ie WAVE or Dedicated Short-Range Communication (DSRC). The final layer is the cloud processing layer which hosts powerful servers. The interconnection in the edge tier is provided by a passive optical network (PON) where the optical network units (ONUs) of the PON connect to the edge nodes and the optical line terminal (OLT) of the PON sources and terminates traffic to the ONUs. The PON infrastructure is a deep PON in this case that links directly to the core network layer leading to the central cloud as shown in Fig. 1.

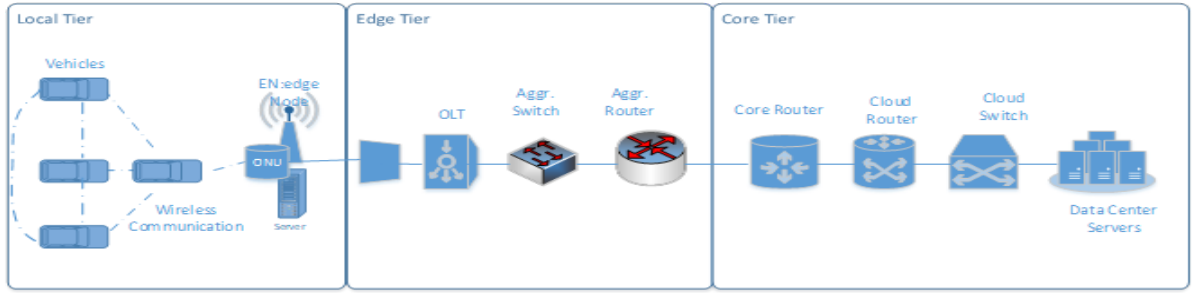

Figure 1: The distributed vehicular, edge and cloud processing architecture

### 3. OPTIMISATION MODEL AND RESULTS

We developed a Mixed Integer linear Programming (MILP) model to optimize the location in which a demand is served in the architecture presented in Figure 1 to minimize power consumption and delay. The power consumption of a processing or network device is composed of the idle power consumed with the device activation; and load-proportional power consumption. For wireless communication interfaces we also consider a distance dependent power consumption for transmission. For the delay, the propagation, transmission and queueing delay are calculated for each source and destination nodes. The model ensures the conservation of processing capacity of each node, as well as the communication interface bandwidth capacity. Also, a processing demand can be served by multiple processing nodes and the full traffic is sent to every processing node regardless of the assigned processing in it. For the queueing delay, we assumed an M/M/1 model, where delay is given as $\frac{1}{\mu - \lambda}$, where μ is the service rate and λ is the arrival rate. However, the number of data packets arriving at a node can vary depending on the model routing decisions, leading to variable arrival rate. A lookup table is used with entries for every possible arrival rate at each node and a precalculated queueing delay associated with it.

The model is evaluated in a parking lot with 8 vehicles parked within 40 × 40 meters space, communicating using DSRC interfaces. The parking lot is served by 2 edge nodes connected to vehicles through WiFi. Each edge node is composed of a server, ONU, and an access point. The edge nodes can also communicate with each other using WiFi. Demands are generated by the vehicles and are composed of two parts, the data to be sent (traffic in kbps), and the processing requirements (in MIPS). Table 1 shows the parameter values for the vehicles and edge nodes. Due to the space limitations, not all values are shown here. More parameters and methods for the parameters estimations are explained in [26]. The modified multi-objective model uses weighting factors to prioritize either the power consumption or the service delay in the resource allocation decision made by the model.

*Table 1: performance evaluation parameters*

| Vehicles Parameters | | Edge Node Parameters | | |
|---|---|---|---|---|
| Parameter | value | Parameter | device | value |
| Processor speed | 800 MHz [27] | Max Power | Raspberry Pi | 12.5 W |
| Max Power | OBU = 10 W [27] | | Access Point | 25 W[28] |
| | WiFi transceiver = 0.612 W [29] | | ONU | 8 W |
| Idle Power | OBU = 5 W [27] | Idle Power | Raspberry Pi | 2 W |
| | WiFi transceiver = 0.000072 W[29] | | Access Point | 5.5 W[28] |
| Tx Power | OBU = +22 dBm [30] | | ONU | 6.8 W (85% of max) |
| | WiFi transceiver = +14 dBm [29] | Tx | Access point | 28 dBm [28] |
| Rx Sensitivity | OBU = -77 dBm [27] | Rx | Access point | -104 dBm [28] |
| | WiFi transceiver = - 72 dBm [29] | Processor Speed | Raspberry Pi | 1.2 GHz |

| WiFi Bandwidth | 150 Mbps [31] | WiFi Bandwidth | Access Point | 150 Mbps [28] |
| --- | --- | --- | --- | --- |
| DSRC Bandwidth | 27 Mbps | Optical Fibre capacity | ONU | 3.75 Gbps |

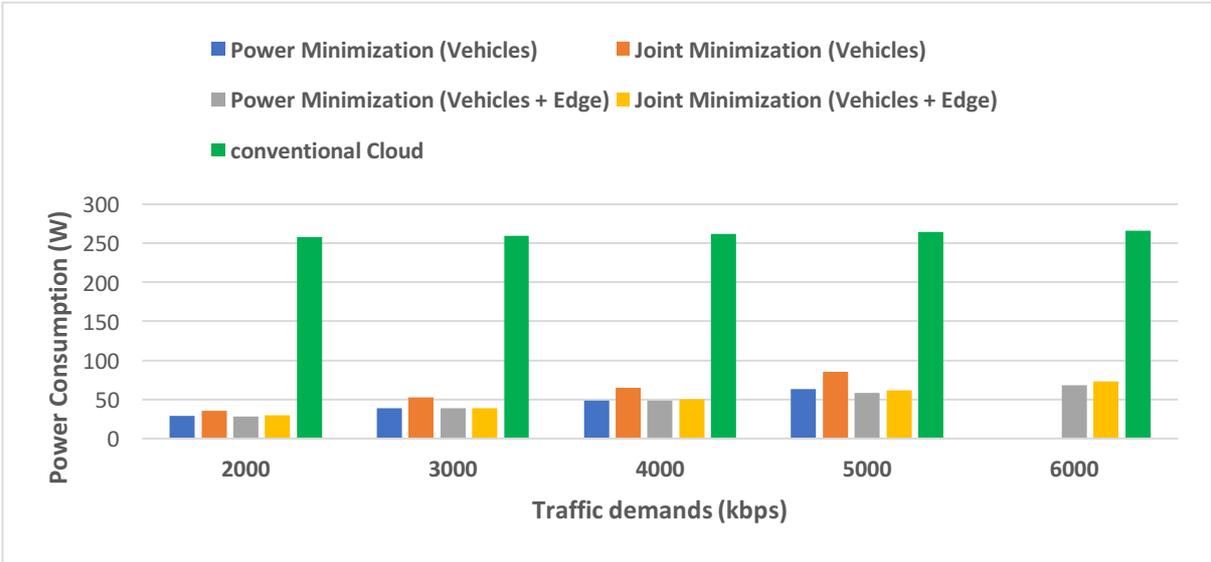

Figure 2 : Total power consumption for each scenario, with varying demand size

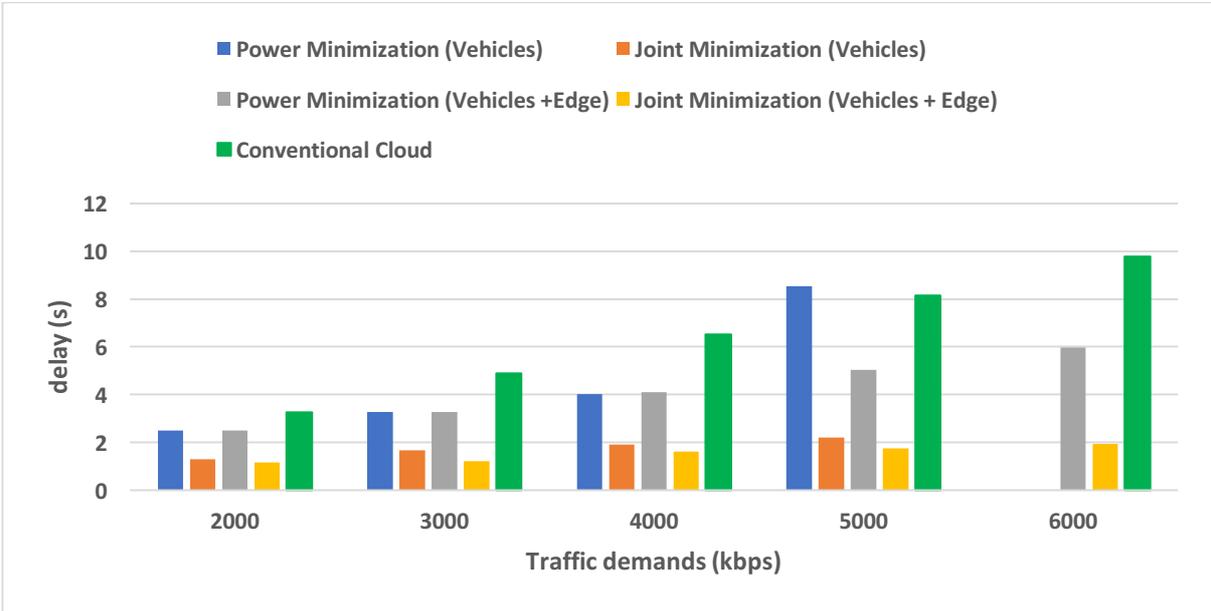

Figure 3:Service Delay for each scenario, with varying demand size

The model is evaluated by varying the demand size of a single request (generated by one vehicle) considering three settings, in the first setting only vehicles have processing capacities, in the second setting both vehicles and edge nodes have processing capacities, and in the third setting conventional cloud is used for processing. For each setting, a scenario is run with an objective to minimize power consumption only. Another scenario is run with an objective to minimize both power and delay with equal weighting factors. We compare the power consumption and delay of each scenario considering the different settings.

Figure 2Figure 2 shows the power consumption versus the traffic demand. It can be noticed that no values are presented for the highest traffic demand (6000 kbps) with the vehicles only setting, as the demand exceeded the vehicles processing capacity. It was established in our work in [26], that vehicles and edge nodes are more efficient for processing in terms of power consumption, and the use of cloud can be limited when demands exceed the processing and communication capacities in the distributed vehicular environment. The same findings hold for this work. However, the power consumption increases by 22%-34% when jointly minimizing delay and power compared to minimizing power only with only vehicles as processing destinations. When using both vehicles and edge nodes to serve the processing demands, enough capacity is available to serve the larger demand of 6000 kbps and the increase in power consumption is limited to 3%-6%. In terms of power saving in comparison with conventional cloud, the power saving drop was more apparent when only vehicles were used for processing. The savings dropped for (89%-76%) range to (86%-68%) range. The case when vehicles and edge nodes were used was more resilient, and the drop is almost negligible, with (89%-73%) ranged maintained.

Figure 3 shows the delay versus the traffic demand. Under distributed processing, the demand can be served in parallel in more than one destination. The delay shown here is the maximum delay experienced by the distributed flows between the source and the different destinations serving the demand. The joint minimization significantly decreases the delay compared to power minimization only. A decrease of 48%-74% can be seen in the vehicles only scenarios, and 54%-67% in the vehicles and edge scenarios. With conventional cloud, the distant location of the cloud from the demand source (around 200-300 km away) increases the propagation delay. Processing in the vehicles and edge nodes reduces the delay by 60-80% compared to the cloud considering the joint minimization.

The increased power consumption and decreased delay can be explained by the fact that optimizing routing to minimize delay results in routing traffic over links with the higher data rate to reduce the transmission delay, which led to choosing WiFi over DSRC which has higher energy per bit leading to increasing in the power consumption.

The results in Figure 2 and Figure 3 show the merits of using edge nodes to support of vehicles in a distributed processing architecture under power only minimization or joint minimization. In addition to the capacity to serve higher demands, the total power consumption and delay are minimized, as the edge nodes have power consumption values close to the vehicles but with higher processing capacity (reference to parameters table). Also, the communication with edge nodes is originally done using WiFi, which has higher data rate, and with the joint optimization relaying upon it more to reduce the delay.

## CONCLUSIONS

In this paper, we studied optimizing resource allocation in a distributed processing architecture based on vehicular and edge computing to jointly minimize power consumption and delay. The results show that the joint minimization significantly decreases the delay while introducing a limited increase in power consumption compared to power minimization only. The results also show the merits of using edge nodes to support vehicles in a distributed processing architecture in terms of higher capacity and reduced power consumption and delay.

## ACKNOWLEDGEMENTS

The authors would like to acknowledge funding from the Engineering and Physical Sciences Research Council (EPSRC), through INTERNET (EP/H040536/1), STAR (EP/K016873/1) and TOWS (EP/S016570/1) projects. All data are provided in full in the results section of this paper.